\documentclass[%
 aip,
rsi,%
 amsmath,amssymb,
 reprint,%
longbibliography
]{revtex4-1}

\usepackage{graphicx}% Include figure files
\usepackage{dcolumn}% Align table columns on decimal point
\usepackage{bm}% bold math
\usepackage{tabularx}
\usepackage{subcaption}

\begin{document}

\preprint{AIP/123-QED}

\title[]{Maximum entropy approach to link prediction in bipartite networks}% Force line breaks with \\

%\title[]{Link prediction for bipartite networks: \\ a maximum entropy approach}% Force line breaks with \\
% \thanks{Footnote to title of article.}

\author{M. Baltakiene}\affiliation{Tampere University of Technology, Tampere, Finland}
\author{K. Baltakys}\affiliation{Tampere University of Technology, Tampere, Finland}
\author{D. Cardamone}\affiliation{GSK Vaccines, Siena, Italy}
\author{F. Parisi}\affiliation{IMT School for Advanced Studies, Lucca, Italy}
\author{T. Radicioni}\affiliation{Scuola Normale Superiore, Pisa, Italy}
\author{M. Torricelli}\affiliation{University of Bologna, Bologna, Italy}\affiliation{ISI Foundation, Turin, Italy}
\author{J. A. van Lidth de Jeude}\affiliation{IMT School for Advanced Studies, Lucca, Italy}
\author{F. Saracco}\affiliation{IMT School for Advanced Studies, Lucca, Italy}

%\homepage{http://www.Second.institution.edu/~Charlie.Author.}
%\affiliation{% Second institution and/or address%\\This line break forced% with \\}%

\date{\today}% It is always \today, today,

\begin{abstract}
%Bipartite networks describe the linking between two classes, as user to items. We consider the problem of link-prediction on such networks, which is used to recover hidden links and predict new links; for example as a recommender system. 
Within network analysis, the analytical maximum entropy framework has been very successful for different tasks as network reconstruction and filtering. %sampling, and randomization. 
In a recent paper, the same framework was used for link-prediction for monopartite networks: link probabilities for all unobserved links in a graph are provided and the most probable links are selected. Here we propose the extension of such an approach to bipartite graphs. We test our method on two real world networks with different topological characteristics. Our performances are compared to state-of-the-art methods, and the results show that our entropy-based approach has a good overall performance. 
%
%Valid PACS numbers may be entered using the \verb+\pacs{#1}+ command.
\end{abstract}

\keywords{Link prediction, Bipartite network, Shannon entropy}%Use showkeys class option if keyword
\maketitle

% \begin{quotation}
% The ``lead paragraph'' is encapsulated with the \LaTeX\ 
% \verb+quotation+ environment and is formatted as a single paragraph before the first section heading. 
% (The \verb+quotation+ environment reverts to its usual meaning after the first sectioning command.) 
% Note that numbered references are allowed in the lead paragraph.
% %
% The lead paragraph will only be found in an article being prepared for the journal \textit{Chaos}.
% \end{quotation}

\section{\label{sec:level1}Introduction}
%:\protect\\ The line
%break was forced \lowercase{via} \textbackslash\textbackslash}
In recent years, the network science has drawn increasing attention in a huge class of real-world phenomena\cite{NEWMAN2010a, caldarelli2010scale-free}, such as financial systems \cite{demasi2006fitness,Battiston2012,Battiston2016,vodenska2017systemic}, brain activity \cite{mastrandrea2017organization, avena2017communication} and socioeconomic systems \cite{Serrano2003, Garlaschelli2004, Fagiolo2009}. Exploring the relations between interconnected objects can lead to a better understanding of the underlying behaviors of those systems. \\ 
Many real-world system can be represented as bipartite networks\cite{guillaume2004bipartite, Latapy2008}, such as collaboration and co-authorships networks\cite{Newman2001}, recommendation networks\cite{Linden2003a,Becatti2018}, financial networks of banks and assets\cite{Gualdi2016a}, biological mutualistic networks\cite{Dormann2009,Suweis2013,Azaele2016a} and trade networks\cite{Hidalgo2009,Cristelli2013}. Standard approaches for their analysis quite often reside on the projection on one of the layers, but nevertheless, the  information contained in the original bipartite network can provide important insights for the comprehension of the phenomena under analysis\cite{Saracco2016}.
\\ 
The study of the network topology is relevant for  many networks processes, such as diffusion phenomena and network resilience. Incomplete or incorrect knowledge over the network topology can cause biases in such analysis. Unfortunately, in real-world networks, the relationships among nodes are not always fully observable, and are subject to frequent changes over time. To overcome these issues, the objective of link prediction is to uncover unobserved or missing connections or forecast the emergence of future relationships from the current topological structure of the network\cite{Liben-Nowell2003,Cannistraci2013,Pan2016}. \\ Link prediction problem is an active research field and many methods have been proposed in the literature. Some methods make use of \textit{local} information, i.e. at node level, while others are based on \emph{global} approaches. In the following we will concentrate on the first class of methods. Also, we can distinguish methods based on similarity measures or likelihood functions. However, only few of the methods proposed in the literature have been applied in the case of bipartite networks\cite{yildirim2014using, Daminelli2015,gao2017}. Among the algorithms which admit bipartite configurations, there are several classes of techiques, such as global and kernel-based methods\cite{Kunegis2009}, extensions of results in monopartite networks to bipartite\cite{Daminelli2015} and projections on the monopartite\cite{gao2017,yildirim2014using}.
\\ \\
\begin{figure}[b]
 \label{fig:bipartite_net} 
 \includegraphics[width=0.5\textwidth]{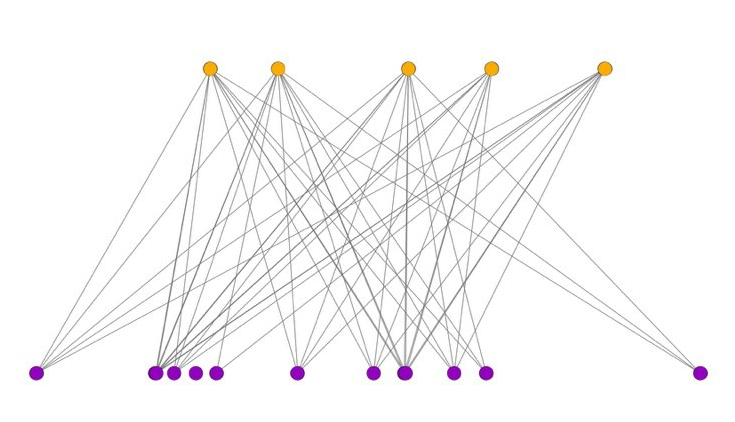}
 \caption{Visualization of a bipartite graph. A subgraph of the \textit{Venezuelan Banks and Assets} bipartite network is shown, with banks on one layer linked to the assets they hold on the other layer. }
\end{figure}
\\ \\
In a recent work, a entropy-based approach was used for link predictions in (monopartite) trade networks, showing good performances\cite{Parisi2018}. This method rests upon the sequential maximizations of Shannon entropy and the network likelihood function, a combination which has been proven to be rather effective both for detecting patterns and to reconstruct the structure of several real-world networks\cite{park2004statistical, Garlaschelli2008,squartini2011analytical}.
In the present paper we are extending this approach to the bipartite case on social and financial networks. As we will see in the following, the entropy-based approach have good performances with respect to available methods, as in the monopartite case\cite{Parisi2018}.
\\ \\
The paper is organized as follows: Section \ref{sec:level3} reports the detailed description of our method, Section \ref{sec:level2} is devoted to the description of the datasets used for the present analysis and Section \ref{sec:level4} illustrates the results which are discussed in Section \ref{sec:level5} where conclusions are also drawn.

\section{\label{sec:level3}Methods}
% \begin{figure*}
%  \caption{Amazon Musical Instruments graph degree frequency.}
%  \label{fig:amazon_mus_inst_deg_freq} 
%  \includegraphics[width=1\textwidth]{amazon_music_instr_degree_freq.eps}
% \end{figure*}
Let us indicate the two layers of the bipartite network as $\top$ and $\bot$; nodes on the layer $\top$ are identified by Latin indices and nodes on the layer $\bot$ with Greek ones. The number of nodes of the two layers is respectively $N_\top$ and $N_\bot$. A bipartite network is described by a biadjacency matrix, i.e. the rectangular matrix $\mathbf{M}_{N_\top\times N_\bot}$ whose entries $m_{i\alpha}$ are 1 if there is an edge connecting $i$ and $\alpha$ and 0 otherwise.
%We will call the total number of links in the bipartite network $E=\sum_{i,\alpha}m_{i\alpha}$.

%We will start to focus on binary and undirected representation of the bipartite network which is the configuration when the matrix entries can assume two values, namely either $m_{i}=1$ or $m_{i}=0$. 
Let us indicate with the symbol $E=\sum_{i,\alpha}m_{i\alpha}$ the corresponding set of observed links and with the symbol $U=N_\top\times N_\bot$ the set of all nodes pair: as a consequence, \textit{U}$\backslash$\textit{E} is the set of non-existent links in the network. In order to study the performance of a link prediction algorithm, the list of edges is usually divided into two separate sets: the training set \textit{E}, used in the "calibration" phase of a given prediction algorithm, and a probe set \textit{E$^{P}$}=\textit{E}$\backslash$\textit{E$^{T}$} which is the set of removed links for testing the algorithm, thus constituting the actual "prediction target". From those definitions, we can indicate with $\mathbf{M^{T}}$ the portion of the adjacency matrix corresponding to the training set. Finally, the union of the missing-links set and the \textit{non-existent links} set \textit{E$^{N}$}$\equiv$\textit{U}$\backslash$\textit{E$^{T}$} will be referred to as to the set of \textit{non-observed links}. 

The following procedure is followed to test our link prediction method and to compare it with alternative algorithms:
\begin{enumerate}
\item the 10\% of links are randomly removed. This operation is repeated 10 times;
\item on each of the reduced matrices we apply the link prediction algorithms;
\item the performance of each algorithm is evaluated by means of different evalutation measures which are then averaged across the 10 iterations.
\end{enumerate}

\subsection{Link prediction methods}\label{sec:lpm}

Link-prediction algorithms output a list of scores to be assigned to non-observed links. The classification algorithms can be divided in two main classes: 
\begin{itemize}
\item \textbf{Similarity-based algorithms} which employ local, quasi-local or global information, such as, respectively, the nodes degree, the degree of common neighbors and the length of paths connecting any two nodes;
\item \textbf{Likelihood-based algorithms} defined by a likelihood function whose maximization provides the probability that any two nodes are connected.
\end{itemize}
The local-based similarity algorithm are based on the fact that the likelihood of an interaction between two non-adjacent nodes is strongly related with mechanisms of organization involving their first and/or second neighbor nodes. Upon indicating with $N(i)$ and $N(\alpha)$ respectively the set of neighbors of $i$ and $\alpha$ and with $N(N(i))$ and $N(N(\alpha))$ respectively the set of the second-order neighbors of the nodes $i$ and $\alpha$, the main similarity indexes are the following:  
\begin{itemize}
\item\textbf{Common neighbors (CN)}: 
\begin{equation}
s^{CN}_{i\alpha}=|(N(i)\cap N(N(\alpha)))\cup(N(\alpha)\cap N(N(i)))|
\end{equation}
is an index counting the neighbors touched by the quadrangles that pass through the nodes $i$ and $\alpha$;
\item\textbf{Resource Allocation (RA)}: $$s^{RA}_{i\alpha}=\sum_{z\in((N(i)\cap N(N(\alpha)))\cup(N(\alpha)\cap N(N(i))))}\frac{1}{|N(z)|}$$ assigns a different weight to the common neighbors of nodes $i$ and $\alpha$ based on its degree;
\item\textbf{Preferential Attachment (PA)}: $$s^{PA}_{i\alpha}=k_i \cdot k_\alpha$$ is simply the degree product of nodes $i$ and $\alpha$, can be used in bipartite networks.
\item\textbf{Cosine Similarity (CS)}: $$ s^{RA}_{i\alpha}= \frac{s^{CN}_{i\alpha}}{\sqrt{|k_i \cdot k_\alpha|}}$$
is based on the Cosine distance between two vectors of same length.
\end{itemize}
In contrast to the existing node-neighborhood-based approaches, the link prediction strategy of other similarity-based models focuses no longer only on groups of common nodes and their node neighbours, but also on the organization of the links between them. In those models, the information content related with the CN nodes is complemented with the topological information emerging from the interactions between them. In order to demonstrate the validity of this theory on several classes of networks, different classical node-based link prediction techniques like CN, JC, RA and PA were reinterpreted. This mathematical reformulation represents the Cannistraci variations\cite{Daminelli2015} of CN, RA and PA respectively renamed Cannistraci-Alanis-Ravasi (CAR), Cannistraci Resource Allocation (CRA) and Cannistraci Preferential Attachment (CPA) and defined in the following way:
\begin{itemize}
\item \textbf{CAR index}: 
\begin{equation*}
s^{CAR}_{i\alpha}=s^{CN}_{i\alpha}\cdot s^{LCL}_{i\alpha}
\end{equation*}
\item \textbf{CRA index}: 
\begin{equation*}
s^{RA}_{i\alpha}=\sum_{z\in((N(i)\cap N(N(\alpha)))\cup(N(\alpha)\cap N(N(i))))}\frac{|\gamma(z)|}{|N(z)|}
\end{equation*}
\item \textbf{CPA index}: 
\begin{equation*}
s^{CPA}_{i\alpha}=e_i \cdot e_{\alpha}+e_i \cdot s^{CAR}_{i\alpha}+e_{\alpha}\cdot s^{CAR}_{i\alpha}+(s^{CAR}_{i\alpha})^2
\end{equation*}
\end{itemize}
where $s^{LCL}_{i\alpha}$ counts the links between the common neighbors of nodes $i$ and $\alpha$, $|\gamma(z)|$ is the number of links of z with the other neighbors of $i$ and $\alpha$, while $e(i)$ and $e(\alpha)$ are the number of external links respectively of nodes $i$ and $\alpha$.

\subsection{The Bipartite Configuration Model approach}\label{sec:BiCM}
In the present paper, the probabilities of the Bipartite Configuration Model\cite{Saracco2015a} (\emph{BiCM}) are used as score function for predicting links, thus extending the approach of \cite{Parisi2018} to bipartite networks.\\
As a first step, an ensemble of (bipartite) graphs is defined with the same amount of nodes per layer. This \emph{ensemble} includes all possible realizations, from the empty network to the fully connected one. 
Then, the objective is to obtain the most general null-model discounting the information of some local constraint. In the Configuration Model, this constraint is represented by the degree sequence, i.e. the number of connections per node. As in Statistical Mechanics, probability per graph can be derived by maximizing the Shannon entropy under the constraint of the degree sequence \cite{Jaynes1957,park2004statistical}. If the constraint is expressed in term of the Hamiltonian $H(G)=\sum_i\theta_i k_i+\sum_\alpha\eta_\alpha k_\alpha$, the probability per graph is:
\begin{equation}\label{eq:prob_per_graph}
P(G|\boldsymbol{\theta}, \boldsymbol{\eta})=\dfrac{e^{-H(G)}}{\mathcal{Z}}\\
\end{equation}
where $\boldsymbol{\theta}$ and $\boldsymbol{\eta}$ are the Lagrangian multiplier respectively of the layer $\top$ and $\bot$ and $\mathcal{Z}$ is the partition function. Interestingly enough, the probability per graph can be factorized in term of probabilities per link:
\begin{equation}\label{eq:prob_per_link}
P(G|\boldsymbol{\theta}, \boldsymbol{\eta})=\prod_{i,\,\alpha}p_{i\alpha}^{m_{i\alpha}}(1-p_{i\alpha})^{1-m_{i\alpha}}
\end{equation}
For further details about the BiCM model, a detailed description is presented in the Appendix \ref{sec:appa}.\\
In order to obtain the actual values of the Lagrangian multipliers, the log-likelihood of the real matrix\cite{Garlaschelli2004,squartini2011analytical} is maximed. It can be shown that it is equivalent to set that the average degree sequence over the ensemble is equal to the one observed in the real matrix:
\begin{equation}\label{eq:log_like}
\begin{split}
\sum_\alpha p_{i\alpha}=&k_i^*\\
\sum_i p_{i\alpha}=&k_\alpha^*.
\end{split}
\end{equation}
The average degree sequence over the ensemble can be expressed in terms of the probabilities per link defined in (\ref{eq:prob_per_link}).

%In fact, it rests upon the information provided by local, topological quantities, which are enforced as constraints of a maximization procedure defined within the Bipartite Configuration Model (BiCM) framework.

\subsection{Evaluation Measures}
\begin{table*}
\caption{\label{tab:table_graph_stats}Data description of the MovieLens graph, and the Venezuelan Banks and Assets graph}
\begin{tabularx}{\textwidth}{XXXXXXXX}
\hline\hline
Graph &	Users (Banks) &	Items & Nodes & Edges & Avg. Degree & Avg. Degree (Users) &	Avg. Degree (Items) \\
\hline\hline
% ADM	& 478235  &	266414 &	744649	& 836006 & 2.25	& 1.75 & 3.14 \\
% AMI	& 339231  &	83046  & 422277 & 500176 & 2.37	& 1.47 & 6.02\\
ML	& 943	  & 1682   & 2625 & 100000 & 76.19 & 106.04 & 59.45\\
VBA\footnote{on average for 103 timestamps} & 45 & 20 & 65 & 912 & 28.06 & 20.27 & 45.60 \\
\end{tabularx}  
\end{table*}
After the link-prediction algorithm has been performed, a number of statistical indices can be used to test its effectiveness. The first index we have considered is the \textit{True Positive Rate} (TPR) (also known with the name of \textit{precision}) which is the percentage of missing-links that are correctly recovered, namely the number $L_m$ of correctly identified missing-links, within the list of the first $|E^P|$ links with the largest score. The TPR is defined as:
\begin{equation}
\makebox{TPR}=\frac{L_m}{|E^P|}
\end{equation}
Another evaluation index is the \textit{area under the ROC curve}, or (AUC). This measures evaluates how many times a method (correctly) assignes a higher score to a missing link with respect to a non existent one. It is formally defined as:
\begin{equation}
\makebox{AUC}= \frac{n'+0.5n''}{n}
\end{equation}
Specifically, for each combination of a missing and non-existent link, if the former scores higher than the latter, the index $n'$ is raised by one unit. If the two links have the same score, $n''$ is raised. The denominator is given by the product of the number of missing links times the number of non-existent ones.
If all scores were i.i.d. the AUC value should be distributed around an expected value of 1/2: therefore, the extent to which the AUC value exceeds 0.5 provides an indication of how much better the algorithm performs than pure chance. Finally, the last index, called \textit{accuracy} (ACC), quantifies the percentage of correctly classified links, namely both the missing ones and the non-existent ones $L_{ne}$, with respect to the total number of non-observed links $|E^N|$:
\begin{equation}
\makebox{ACC}=\frac{L_m+L_{ne}}{|E^N|}
\end{equation}

\section{\label{sec:level2}Data}
The following datasets have been employed to test the link-prediction method:
\begin{itemize}
% \item \textbf{Amazon Digital Movies (ADM)}: Amazon dataset involving digital movies ratings. For all of them, the ratings range for each purchased product span from 1 to 5;
% \item \textbf{Amazon Musical Instruments (AMI)}: Amazon dataset involving musical instruments ratings. As before, for all of them, the ratings range for each purchased product span from 1 to 5;
\item \textbf{MovieLens (ML)}: MovieLens\cite{MovieLens2018} datasets were collected by the GroupLens Research Project at the University of Minnesota. This data set consists of 100000 ratings (1-5) from 943 users on 1682 movies. Each user has rated at least 20 movies and is characterized by some demographic information, such as age, job, sex, state and zipcode. The data was collected through the MovieLens web site (\url{movielens.umn.edu}) during the seven-month period from September 19th, 1997 through April 22nd, 1998. For the set of movies, there is information on the release year, title and genre. Each user can review a movie with a score that ranges from 1 to 5, according to his level of appreciation. We binarize the network by drawing an edge for a user-movie pair if the user has reviewed the movie;
\item \textbf{Venezuelan Banks and Assets (VBA)}: Bipartite networks of positions that 69 Venezuelan banks hold in 20 asset classes in the period between December 2013 and June 2015. The dataset was firstly presented and analyzed in \cite{Levy-Carciente2015}. The binarized network has an edge between a bank-asset pair, if the position the bank held in the asset class has a value greater than zero at a given timestamp.
\end{itemize}
The generic statistics of the graphs produced from the datasets that were used in this analysis are provided in Table \ref{tab:table_graph_stats}.

\section{\label{sec:level4}Results}

\begin{figure}
 \includegraphics[width=0.5\textwidth]{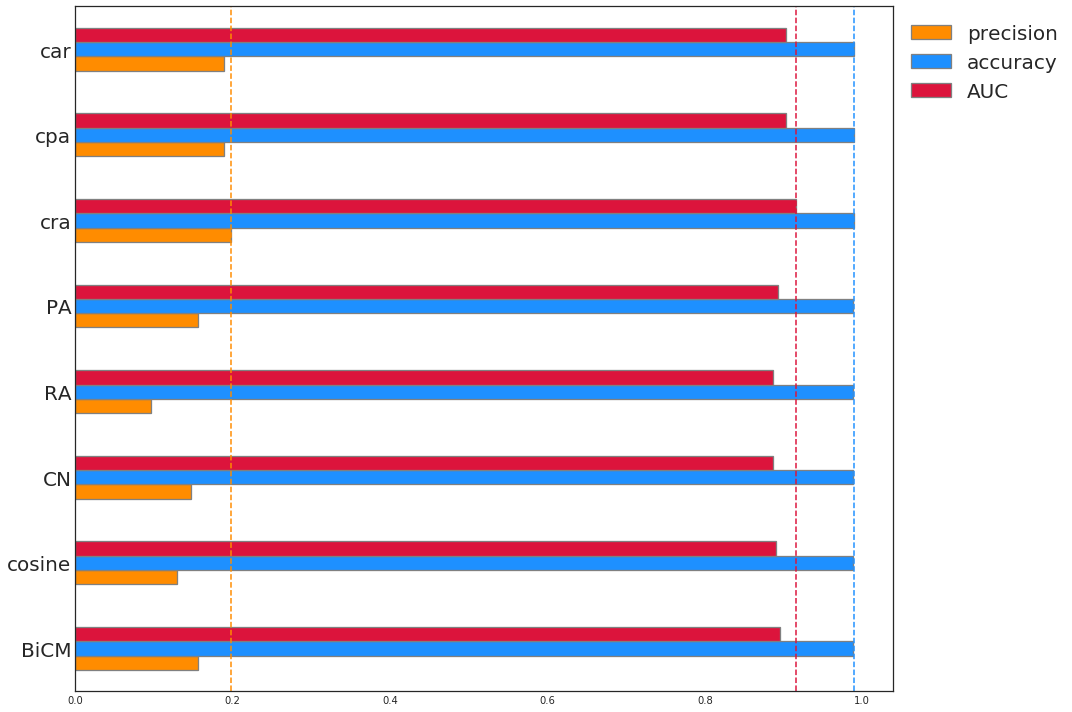}
 \caption{Link prediction performance on the MovieLens dataset, a bipartite graph of users linked to movies they reviewed. The BiCM method performs on par with the alternative methods on all evaluation measures. \label{fig:movielens_prediction_performance}}
\end{figure}

\begin{figure}
\centering
  \begin{subfigure}[b]{0.5\textwidth}
  \caption{Precision \label{fig:bank_precision_performance}}
  \includegraphics[width=1\textwidth]{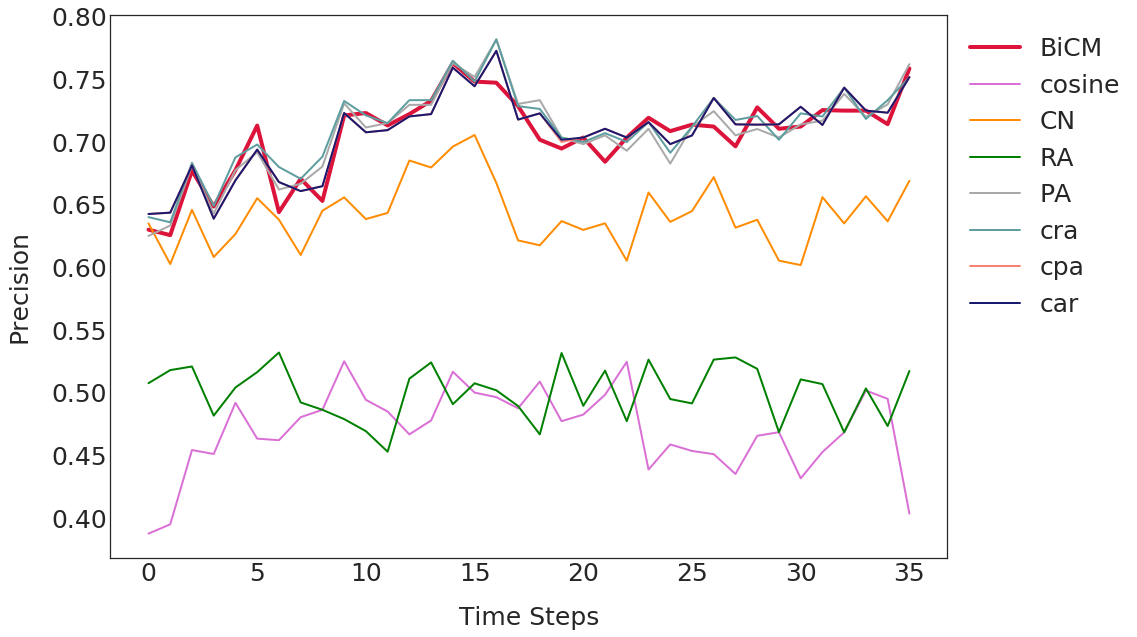}
  \end{subfigure}

  \begin{subfigure}[b]{0.5\textwidth}
  \centering
  \caption{Accuracy \label{fig:bank_accuracy_performance} }
  \includegraphics[width=1\textwidth]{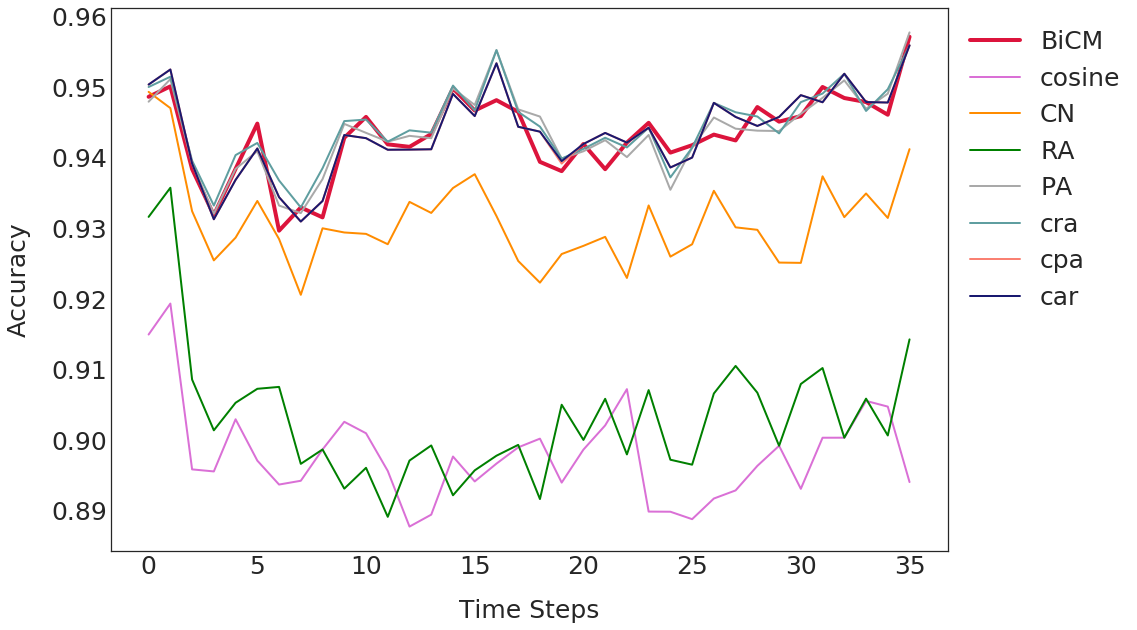}
  \end{subfigure}

  \begin{subfigure}[b]{0.5\textwidth}
  \caption{AUC \label{fig:bank_auc_performance} }
  \includegraphics[width=1\textwidth]{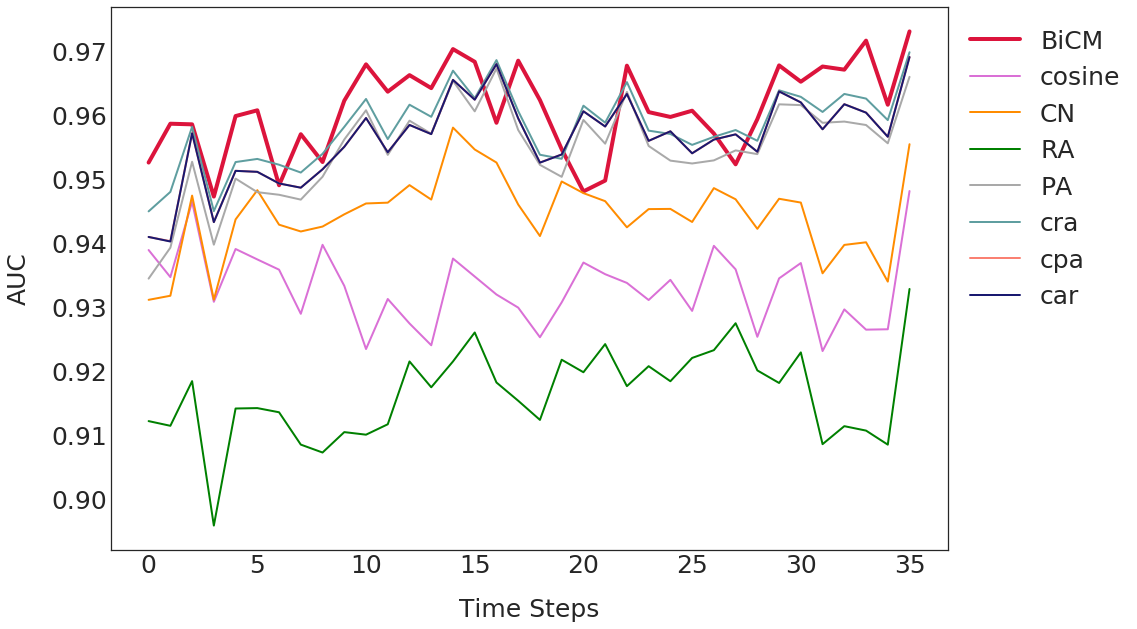}
  \end{subfigure}
\caption{Link prediction performance on the Venezuelan Banks and Assets dataset, a bipartite graph which describes the types of assets in which banks held a position at various time-snapshots between December 2013 and June 2015. Performance is measured by Accuracy, Precision and AUC. The BiCM method is among the best performing methods in all time snapshots, and has best performance in some of these.}
\end{figure}

The results of the algorithm performances are presented in Table \ref{tab:table_movielens_results} and Figure \ref{fig:movielens_prediction_performance} for the MovieLens data set, and the metrics comparison for the Venezuelan Banks and Assets are in the Figures \ref{fig:bank_precision_performance}, \ref{fig:bank_accuracy_performance} and \ref{fig:bank_auc_performance}.\\
The results of the link prediction in bipartite networks are averaged over 10 iterations for each method, with exception for the BiCM method for the MovieLens data set the average is taken from 7 iterations. Table \ref{tab:table_movielens_results} shows the averaged measure and the standard deviation (SD).\\
The results for our entropy based algorithm (BiCM) are comparable and show strong performance as opposed to the benchmark algorithms. For the MovieLens network BiCM comes in third place for the accuracy (ACC) measure and is the fourth best algorithm for the precision (TPR) and AUC (see Table \ref{tab:table_movielens_results}) and is closely trailing the best performers. For the Venezuelan Banks and Assets networks, BiCM link prediction algorithm is among the leaders for the precision and accuracy measures (see Figure \ref{fig:bank_precision_performance} for multiple time periods). Our method shows strong competition with the other leading algorithms and is considerably stronger than RA or cosine similarity methods' performances.
Furthermore, inspecting the Figure \ref{fig:bank_auc_performance} it can be observed that BiCM method dominates the others in AUC measure. 
For all algorithms in the Venezuelan Banks and Assets network analysis, precision ranges between 0.3875 and 0.7818, accuracy is between 0.8877 and 0.9577, and  AUC values change from 0.8958 to 0.9730.
All algorithms perform better on the Venezuelan Banks and Assets than on the MovieLens data set with respect to precision measure, i.e. the rate of true positive values.

\begin{table}
\caption{\label{tab:table_movielens_results} Results: MovieLens performance comparison}
\begin{tabular}{llll}
\hline \hline
Method	&	ACC (SD) &	TPR (SD)	& AUC (SD) \\
\hline \hline
BiCM	& \textbf{0.98873} (0.00003) &	\textbf{0.15658} (0.002)	& \textbf{0.8946}  (0.002) \\
cosine	& 0.98836 (0.00005) &	0.12905 (0.004)	& 0.8903  (0.0007) \\
car	 &	0.98917 (0.00003)	&  0.18975 (0.003) & 0.9028  (0.001)\\
CN	 &	0.98860 (0.00003)	&   0.14713 (0.002) &  0.8868 (0.0009)\\
cpa	 &	0.98917 (0.00003)	& 0.18975  (0.003) & 0.9028  (0.001)\\
cra	 &	0.98929 (0.00003)	& 0.19856  (0.002) &  0.9163 (0.001)\\
PA   &	0.98873 (0.00003)	&  0.15672 (0.002) &  0.8932 (0.001)\\
RA	&	0.98793 (0.00004)	&  0.09712 (0.003) &  0.8863 (0.0007)\\
\end{tabular}
\end{table}

%\begin{table}
%\caption{\label{tab:table_banks_results} Results: Venezuelan Banks and Assets metrics comparisons}
%\begin{tabular}{cccc}
%\hline\hline
%Method	&	ACC (SD) &	TPR (SD)	& AUC (SD) \\
%    \hline\hline
%BiCM	& 0.94 (0.01) &	 0.70 (0.06)	&   0.96 (0.02) \\
%CS	& 0.90 (0.01) &	 0.47 (0.06)	&  0.93 (0.02) \\
%\end{tabular}
%\end{table}

\section{\label{sec:level5}Conclusion and Discussion}
Link-prediction is a method that can be leveraged for a wide array of tasks, as compensating for missing information\cite{Liben-Nowell2003}. Recently, it was proposed to employ entropy based null-model probabilities \cite{park2004statistical,Garlaschelli2004,squartini2011analytical}
as score function for predicting links\cite{Parisi2018}: missing links with high probability are likely to be present. In the present paper we extend this approach to bipartite networks, thus using as score function the probability of the bipartite configuration model\cite{Saracco2015a}.\\ In order to test our predictions, we first randomly remove a fraction of the links present in the real network and then use our procedure to predict the same amount of links. On the real world bipartite networks of user-movie ratings and bank-asset positions, we compared the performances of our proposed method to seven alternative local information based methods. On all datasets and all evaluation measures our method is able to consistently predict missing links.

It is not surprising that our approach has better performances on financial data, than on the social network of Movielens\cite{Harper2015,MovieLens2018}: indeed in the latter case it is known that a collaborative filtering recommendation system\cite{Resnick1994b} was employed\cite{Harper2015}. Nevertheless, our results have similar performances with other known methods. Moreover, it is remarkable that our approach, that is based on local constraints has performances of the same order of quasi-local methods as the bipartite extension of Cannistraci corrected scores\cite{Cannistraci2013, Daminelli2015}. For financial networks, for which the bipartite configuration model is known for having good performances, results are more promising: the BiCM induced link prediction has almost always the best performances. 
\\ \\
Our method can be naturally extended to (bipartite) review networks, as the ones in which a users can give a rating to a certain item. While the prediction of  both the existence of links and their strength is not trivial, the recent extension to bipartite score network of the configuration model\cite{Becatti2018} makes the task more promising, thus overcoming non-trivial extensions of the present algorithms\cite{Cannistraci2013, Daminelli2015}.

\begin{acknowledgments}
This work is the output of the Complexity72h workshop (\url{https://complexity72h.weebly.com/}), 
held at IMT School for Advanced Studies in Lucca, 7-11 May 2018.\\ F.P., J.v.L.d.J, F.S. thank Tiziano Squartini for interesting discussions, Carolina Becatti and Giuseppe Trapani for help with codes.
K.B. is funded from the European Union’s Horizon 2020 research and innovation programme under the Marie Sk\l{}odowska-Curie grant agreement No 675044 (BigDataFinance). F.S. was supported by the EU projects CoeGSS (Grant No. 676547), MULTIPLEX (Grant No. 317532), Openmaker (Grant No. 687941), SoBigData (Grant No. 654024), and the FET projects SIMPOL (Grant No. 610704), DOLFINS (Grant No. 640772).
\end{acknowledgments}

\appendix

\section{Bipartite Configuration Model derivation}\label{sec:appa}
Let $\mathcal{G}$ be the set of all possible bipartite graphs with respectively $N_\top$ nodes on the upper layer and $N_\bot$ on the lower one. We can define the entropy over this set as:
\begin{equation}\label{eq:entropy}
\mathcal{S}=-\sum_{G\in\mathcal{G}} P(G)\ln P(G).
\end{equation}
Let us maximize the entropy in Equation (\ref{eq:entropy}) by fixing the average value of the degree sequence. Introducing Lagrangian multipliers $\boldsymbol{\theta}$ and $\boldsymbol{\eta}$ for the topological constraints and $\boldsymbol{\alpha}_0$ for the normalization of the probability, the constrained entropy maximization turns into maximizing the function:
\begin{equation}
\begin{split}
\tilde{\mathcal{S}}=-&\sum_{G\in\mathcal{G}} P(G)\ln P(G)\\
+&\sum_i \theta_i(k_i^*-\langle k_i\rangle)\\
+&\sum_\alpha \eta_\alpha(k_\alpha^*-\langle k_\alpha\rangle)\\
+&\boldsymbol{\alpha}_0(1-\sum_{G\in\mathcal{G}}P(G)),
\end{split}
\end{equation}
with respect to the probability per graph.Introducing the Hamiltonian as in Section \ref{sec:BiCM}, the maximization of the entropy returns a probability per graph as in Equation (\ref{eq:prob_per_graph}):
\begin{equation}
\begin{split}
P(G|\boldsymbol{\theta},\boldsymbol{\eta})=&\dfrac{e^-{H(G)}}{\mathcal{Z}}\\
=&\prod_{i,\alpha}\dfrac{(x_iy_\alpha)^{m_{i\alpha}}}{1+x_iy_\alpha},
\end{split}
\end{equation}
where $x_i=e^{-\theta_i}$ and $y_\alpha=e^{-\eta_\alpha}$. Following the lines of Section \ref{sec:BiCM}, we can interpret $\dfrac{x_iy_\alpha}{1+x_iy_\alpha}$ as (independent) probabilities per links. In order to obtain the actual values of the Lagrangian multipliers, let us maximize the log-likelihood $\mathcal{L}$, defined as: 
\begin{equation*}
\begin{split}
\mathcal{L}=&\ln P(G|\mathbf{x}, \mathbf{y})\\
=&\sum_{i,\alpha}m_{i\alpha}\ln(x_iy_\alpha)-\ln(1+x_iy_\alpha),
\end{split}
\end{equation*}
whose maximization on the real network returns exactly the conditions reported in Equation (\ref{eq:log_like}).
\section{Other link prediction methods in terms of quantities per node}
In the present section, the score functions introduced in section \ref{sec:lpm} are rewrited in terms of the biadjacency matrix. The rationale is to provide a consistent formal framework in which all quantities can be expressed. Let us start with the CN: the bipartite extension of the Common Neighbors\cite{Daminelli2015} is the number of nodes in the subgraph defined by the first neighbors of nodes ($i$,$\alpha$). In other words, the Common Neighbors counts the number of nodes involved in at least one ``quadrangular''\cite{Daminelli2015} (or \emph{X-motif}\cite{Saracco2015a}) if $i$ and $\alpha$ were present. This can be expressed as: 
\begin{equation}\label{eq:sCN_adj}
\begin{split}
s^{CN}_{i\alpha}=&\sum_j m_{j\alpha}\Theta_\text{Heaviside}(\sum_\beta m_{i\beta}m_{j\beta})\\
+&\sum_\beta m_{i\beta}\Theta_\text{Heaviside}(\sum_j m_{j\alpha}m_{j\beta}),
\end{split}
\end{equation}
where $\Theta_\text{Heaviside}$ is the \textit{Heaviside Theta} which has a value equal to 1 if its argument is positive and to 0 otherwise. The first term in Equation (\ref{eq:sCN_adj}) considers the number of nodes of the layer $\top$ that are involved in, at least, one quadrangular insisting on ($i$,$\alpha$), the second is the analogous for the layer $\bot$. It is important to notice which it is not necessary to set $j\neq i$ or $\beta\neq\alpha$ in the summations since $m_{i\alpha}=0$.\\ 
In the Resource Allocation, each of the terms contributing to $s_{i\alpha}^{CN}$ is weighted by the inverse of its degree, thus:
\begin{equation}\label{eq:sCN_adj}
\begin{split}
s^{RA}_{i\alpha}=&\sum_j \dfrac{m_{j\alpha}}{k_j}\Theta_\text{Heaviside}(\sum_\beta m_{i\beta}m_{j\beta})\\
+&\sum_\beta \dfrac{m_{i\beta}}{k_\beta}\Theta_\text{Heaviside}(\sum_j m_{j\alpha}m_{j\beta}),
\end{split}
\end{equation}
where we implicitly consider $m_{j\alpha}/k_j=0$ if both the numerator and the denominator are 0.\\

In this formal framework, the expression of Local Community Links (LCL) is much simpler, since it is defined as the number of links in the subgraph defined by the neighbors of $(i,\alpha)$. In fact, it is the number of quadrangular that would be closed by the presence of the link $(i,\alpha)$, i.e. 
\begin{equation}
s^{LCL}_{i\alpha}=\sum_{j,\beta} m_{j\alpha} m_{i\beta}m_{j\beta}.
\end{equation}
By construction, the value of the (bipartite) $s^{LCL}_{i\alpha}$ is limited from below by $s_{i\alpha}^{CN}$. The quantity $\gamma$\cite{Daminelli2015} defined in Section \ref{sec:level3} represents the degree in the subgraph of the neighbors of $(i,\alpha)$. For a given node $j\in\top$, $\gamma(j)$, it can expressed as:
\begin{equation*}
\gamma(j)=m_{j\alpha} \sum_{\beta} m_{i\beta}m_{j\beta}
\end{equation*}
while the expression for a generic node $\beta\in\bot$ is analogous.
\section*{References}
\nocite{*}
\bibliography{72h_complexity}% Produces the bibliography via BibTeX.

\end{document}